\documentclass[prl,superscriptaddress,floatfix,twocolumn,amsmath]{revtex4-2}
\usepackage{graphicx}
\usepackage{wasysym}
\usepackage{hyperref}
\usepackage{color}
\def\newr{\color{black}}
\def\newb{\color{black}}

\def\lco{La$_2$CuO$_4$}

\def\lsco{La$_{2-x}$Sr$_x$CuO$_4$}
\def\lbco{La$_{2-x}$Ba$_x$CuO$_4$}

\def\ybco{YBa$_2$Cu$_3$O$_{6+x}$}

\def\bscco{Bi$_2$Sr$_2$CaCu$_2$O$_{8+\delta}$}

\begin{document}

\title{Experimental Evidence that Zn Impurities Pin Pair-Density-Wave Order in La$_{2-x}$Ba$_x$CuO$_4$}

\author{P. M. Lozano}
\affiliation{Condensed Matter Physics \&\ Materials Science Division, Brookhaven National Laboratory, Upton, New York 11973-5000, USA}
\affiliation{Department of Physics and Astronomy, Stony Brook University, Stony Brook, NY 11794-3800, USA}
\author{G. D. Gu}
\author{J. M. Tranquada}
\email[Corresponding author: ]{jtran@bnl.gov}
\affiliation{Condensed Matter Physics \&\ Materials Science Division, Brookhaven National Laboratory, Upton, New York 11973-5000, USA}
\author{Qiang Li}
\email[Corresponding author: ]{qiangli@stonybrook.edu}
\affiliation{Condensed Matter Physics \&\ Materials Science Division, Brookhaven National Laboratory, Upton, New York 11973-5000, USA}
\affiliation{Department of Physics and Astronomy, Stony Brook University, Stony Brook, NY 11794-3800, USA}

\date{\today}
\begin{abstract}
Both Zn-doping and $c$-axis magnetic fields have been observed to increase the spin stripe order in \lbco\ with $x$ close to 1/8.  For $x=0.095$, the applied magnetic field also causes superconducting layers to decouple, presumably by favoring pair-density-wave order that consequently frustrates interlayer Josephson coupling.  Here we show that introducing 1\%\ Zn also leads to an initial onset of two-dimensional (2D) superconductivity, followed by 3D superconductivity at lower temperatures, even in zero field.  We infer that the Zn pins pair-density-wave order locally, establishing the generality of such behavior.  
\end{abstract}
\maketitle

Among cuprate high-temperature superconductors, spin-stripe order is seen most clearly in the \lco-based family \cite{fuji12a}.  This order, occurring together with charge-stripe order, is strongest for doped hole concentration, $x$, near 1/8.  It is significantly enhanced in those compositions with the low-temperature-tetragonal crystal structure \cite{tran97a,ma20,klau00}, where the Cu-O bond-lengths are anisotropic;  a prominent example is \lbco\ (LBCO) \cite{fuji04,huck11}.  Spin-stripe order can be strongly enhanced in \lsco\ (LSCO) for similar $x$ by application of a magnetic field $H$ perpendicular to the CuO$_2$ planes  \cite{lake02,khay05,chan08} or by Zn doping \cite{hiro01,gugu17}.  In the cases of LBCO and Nd- and Eu-doped LSCO with $x\sim1/8$, the onset of spin-stripe order is associated with the occurrence of two-dimensional superconductivity (2D SC) \cite{li07,ding08,shi20b}.  The depression of 3D SC order due to the frustration of the interlayer Josephson coupling is attributed to pair-density-wave (PDW) order \cite{berg07,agte20}.  

In this letter, we test whether such decoupling can also be induced by Zn doping.  At first glance, this might seem doubtful.  It has long been known that substitution of a small concentration of Zn for Cu in cuprate superconductors such as LSCO severely depresses the bulk superconducting ordering temperature, $T_c$ \cite{koik92,fuku96}.  For $x\sim0.12$, it can take as little as 2\%\ Zn to quench the bulk superconductivity \cite{adac04}.

At the same time, the impact of Zn and $H$ on the electronic properties of LSCO is strikingly similar.  For example, an applied $H$ (above the lower critical field) penetrates the sample as vortices, with a suppression of superfluid density within the vortex cores.  From muon spin rotation studies, a similar effect was inferred for Zn-doped samples, with the reduction in superfluid density per Zn interpreted with the ``Swiss cheese'' model, where each Zn impurity eliminates superfluid density of a CuO$_2$ within an area comparable in size to a magnetic vortex core \cite{nach96}.  For underdoped LSCO, both Zn and $H$ can cause a low temperature upturn of the in-plane resistivity, where the onset temperature of the upturn is a maximum for $x\sim0.12$ \cite{komi04}.  

It is usually assumed that the impact of $H$ and Zn is to kill electron pairing in their vicinity; however, a recent scanning tunneling microscopy (STM) study of \bscco\ has provided evidence for PDW order in the halo region surrounding magnetic vortex cores \cite{edki19}.  Given the other similarities in responses to $H$ and Zn, could it be that Zn locally stabilizes PDW order?  If this is the case, then we would expect to see the introduction of Zn into a 3D SC result in decoupled superconducting layers for some range of temperature above the onset of 3D superconductivity.

Our model system is LBCO with $x=0.095$, where the $T_c$ for 3D order is 32~K in zero field; application of $H$ along the $c$ axis decouples the superconducting planes \cite{wen12b,steg13}.  With substitution of 1\%\ Zn for Cu, the bulk $T_c$ is reduced to $\sim17$~K \cite{wen12a}.  Here, we use anisotropic resistivity and susceptibility measurements to show that 2D SC appears below 26~K, well above the 3D SC transition, and consistent with the idea that Zn impurities locally pin patches of PDW order.  As we will discuss, our results have interesting implications for the impact of defects on the energy balance between uniform d-wave and PDW orders and for the quasiparticle-interference analysis of STM data.

The single crystals studied here were grown by the floating-zone technique \cite{huck11b}. The samples are cut from the same crystal boule used previously \cite{wen12a,steg13}. Meticulous care has been taken in cutting, orienting, and polishing the crystals for transport measurements to keep any misalignment between the direction of current flow and major crystalline axes to less than $1^\circ$.  

\begin{figure}[t]
\centerline{\includegraphics[width=0.9\columnwidth]{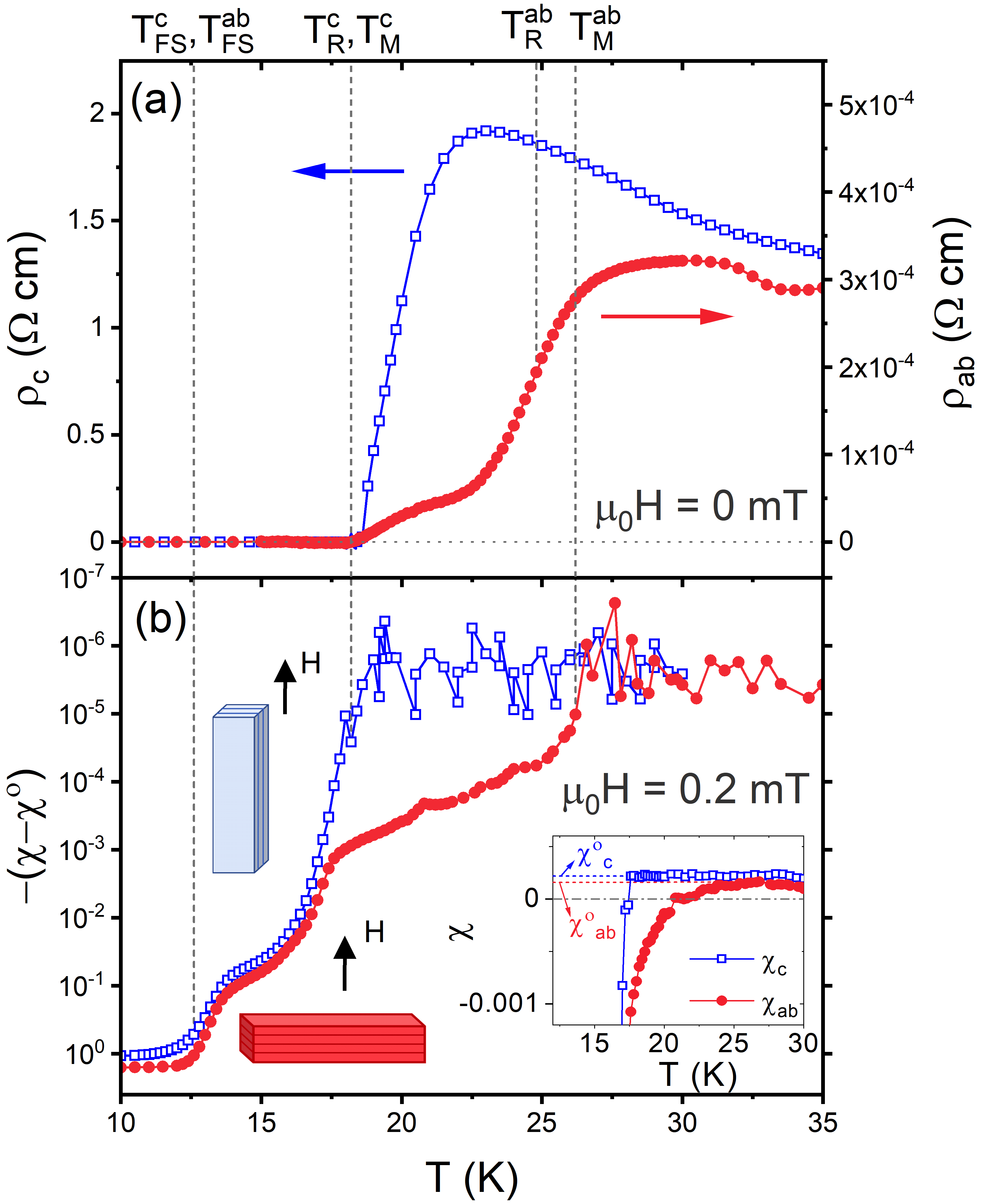}}
\caption{(a) Temperature dependence of the resistivity (in zero applied magnetic field) for current along the $c$ axis (blue squares) and parallel to the $ab$ planes (red circles). (b) Temperature dependence of the volume magnetic susceptibility with a field of $\mu_0H=0.2$~mT applied parallel to the planes, $\chi_c$ (blue squares), and perpendicular to the planes, $\chi_{ab}$ (red circles). {\newr The inset shows the susceptibilities on a linear scale near the initial transitions; linear fits to the normal-state, and extrapolated to low $T$ are indicated by dashed lines.  Subtraction of these fits enables the logarithmic-scale plots of the main panel.}
%Note that we take the absolute magnitude of $\chi$, which is negative, so that we can plot it on a logarithmic scale.  
Characteristic temperatures discussed in the text are indicated by vertical dashed lines with labels at the top edge. }
\label{fg:ZF} 
\end{figure}

\begin{figure*}
\centerline{\includegraphics[width=1.6\columnwidth]{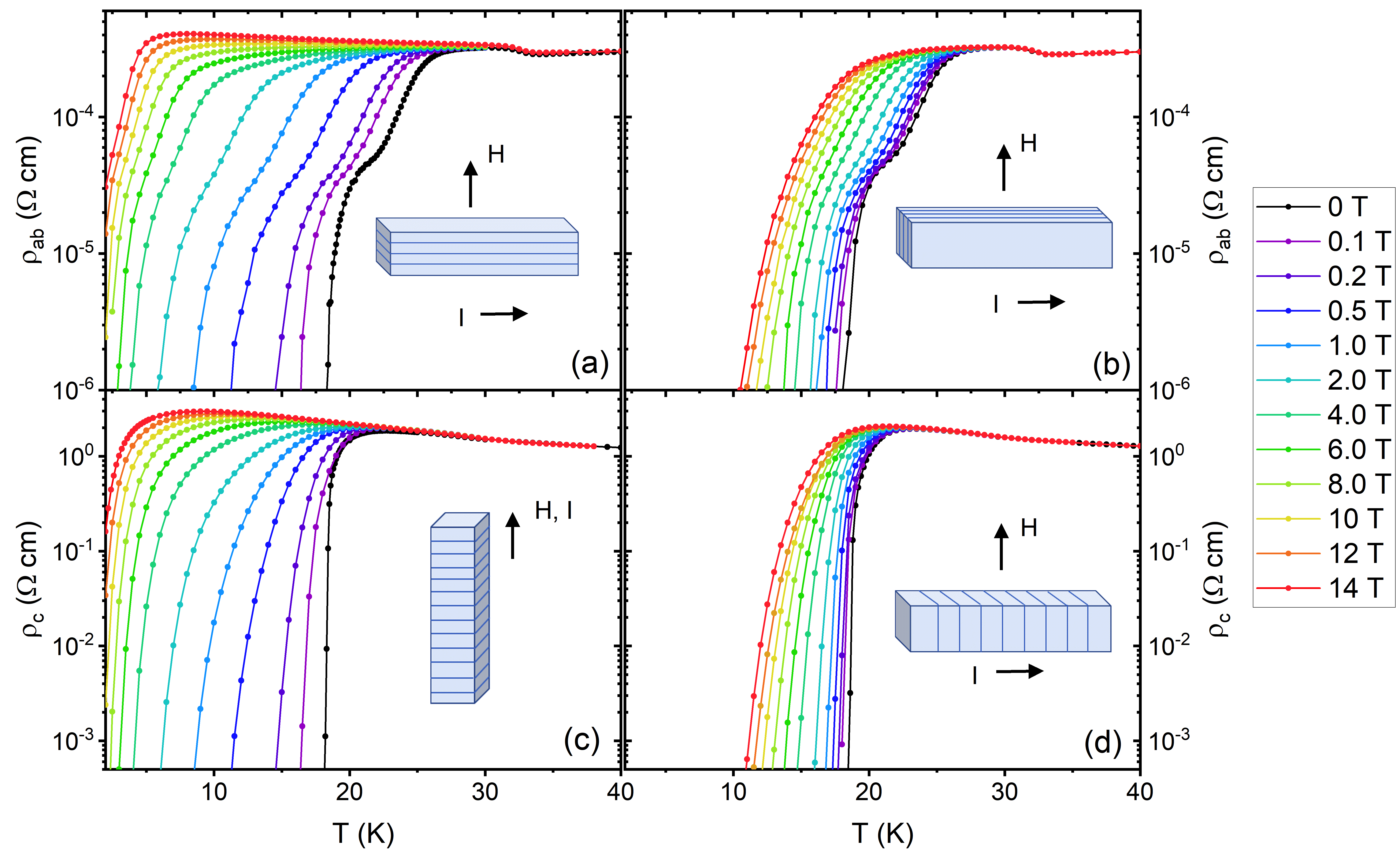}}
\caption{Semi-log plots of the resistivity vs.\ temperature for two field directions:  $\rho_{ab}$ with (a) ${\bf H} \parallel {\bf c}$ and (b) ${\bf H}\perp {\bf c}$;  $\rho_c$ with (c) ${\bf H} \parallel {\bf c}$ and (d) ${\bf H}\perp {\bf c}$.  The value of $\mu_0H$ for each color-coded curve is indicated in the legend on the right.}
\label{fg:rho} 
\end{figure*}

Transport measurements were carried out by the four-probe in-line method on four crystals for $ab$-plane resistivity, $\rho_{ab}$, and two crystals for $c$-axis resistivity, $\rho_c$, in a 14 T Quantum Design Physical Properties Measurement System. To measure $\rho_{ab}$ and $\rho_c$, current contacts were made at the ends of each crystal along the long direction (typical dimensions: $\sim4$--7~mm long, 0.2--1.6~mm$^2$ cross-sectional area) to ensure a uniform current flow throughout the entire sample; voltage contacts were made in direct contact with the $ab$-plane edges \cite{SM}. All samples show similar temperature and magnetic field dependence for the same orientation of current with respect to the desired crystal axis. The reproducibility for both $\rho_{ab}$ and $\rho_c$ are presented in \cite{SM}.  The volume magnetic susceptibility (defined as $\chi= M/\mu_0 H$, where $M$ is the volume magnetization in Tesla, $\mu_0 H$ is the external magnetic field in Tesla) on a crystal from the same slab was measured in a Quantum Design Magnetic Properties Measurement System with a SQUID (superconducting quantum interference device) magnetometer. 100\%\ magnetic shielding was observed at temperatures below 10~K and low fields.

Let us first consider the superconducting transition. Figure~\ref{fg:ZF}(a) shows the temperature dependence of $\rho_{ab}$ and $\rho_c$ in zero field. $\rho_{ab}$ shows metallic behavior (data up to 300 K are presented in \cite{SM}) above the superconducting transition, with a distinct elevation at  32.5~K in association with a first-order structural transition  \cite{wen12a}.  The onset of in-plane superconductivity occurs at $\sim  26.5$~K, where $\rho_{ab}$ begins to  decrease below the normal-state metallic behavior; this also corresponds to the initial rise of spin-stripe order measured by neutron diffraction \cite{wen12a}.  To provide a more robust measure of the resistive onset of 2D superconductivity, we define $T_R^{ab}$ as the maximum of $d\rho_{ab}(T)/dT$, which is at 25.4~K.  On further cooling, $\rho_{ab}$ exhibits a hump before gradually dropping to an unmeasurably small value.  The magnitude of this hump varies among samples, and hence we believe that it is due primarily to a small contribution from $\rho_c$ resulting from a slight misorientation of the crystal faces relative to the principle axes.  For example, from the measured ratio of $\rho_c/\rho_{ab} \sim 4\times10^6$ at 22.5~K (where the maximum in $\rho_c$ occurs), the finite hump would be explained by a misorientation angle of $\sim0.29^\circ$.  To estimate the 3D resistive superconducting transition, we use the electric field criterion of $E = 1\times10^{-5}$~volt/cm and find from $\rho_c$ that $T_{\rm R}^c\approx17.5$~K.

Figure~\ref{fg:ZF}(b) shows the zero-field-cooled measurement of $\chi$ on one crystal for two orientations of the external magnetic field of 0.2~mT.   With the field applied along the $c$ axis, the shielding supercurrent flows within the $ab$ planes and we label it $\chi_{ab}$; with the field parallel to the planes, the shielding current is limited by its component along $c$, so it is labelled $\chi_c$.  From $\chi_{ab}$, the onset of {\newr a diamagnetic decrease} occurs at $T_M^{ab} = 26.2$~K, similar to $T_{\rm R}^{ab}$; {\newr absolute diamagnetism is established below 20.5~K}.  No detectable diamagnetism in $\chi_c$ (beyond the noise level of  $5\times10^{-6}$) was found until the temperature was reduced below $T_{\rm M}^c\approx18$~K, indicating onset of Josephson coupling between the superconducting planes, consistent with $T_{\rm R}^c$.  Now, the response below this transition only reaches about 10\%\ of full shielding.  Another transition occurs at $T_{FS}^{ab}\approx 12.5$~K (and, equivalently, $T_{FS}^c$) that leads to full shielding at low temperature.

Next, we consider the field dependence of the resistivities. As shown in Fig.~\ref{fg:rho}(a) and (c), even a modest $c$-axis magnetic field causes both the temperature of the initial drop in $\rho_{ab}$, corresponding to $T_{\rm R}^{ab}$, and the approach of $\rho_{c}$ to zero, at $T_{\rm R}^c$, to decrease fairly rapidly, although they remain distinct.  (Here, by plotting the log of the resistivities, we can see that both $\rho_{ab}$ and $\rho_c$ approach zero at very similar temperatures.)  When the field is applied parallel to the planes (and perpendicular to $c$), as in Fig.~\ref{fg:rho}(b) and (d), the transition temperatures decrease much more slowly.  This anisotropic response is consistent with the previous observations on the Zn-free sample \cite{wen12b}.  When the field is parallel to the planes, it has minimal impact on the superconductivity within the planes, which is the dominant factor in determining the temperature at which the resistivities approach zero.

\begin{figure}[t]
\centerline{\includegraphics[width=0.9\columnwidth]{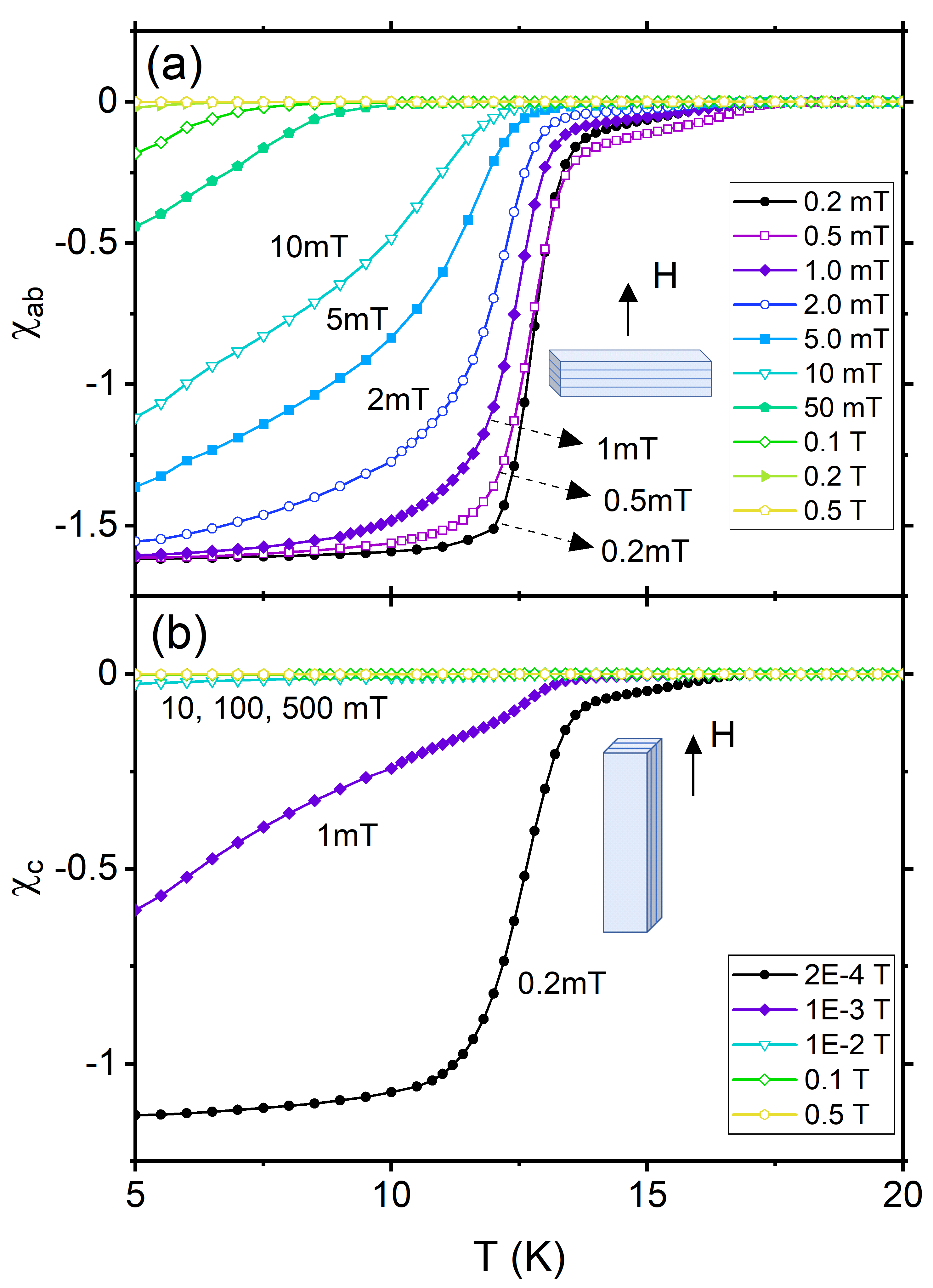}}
\caption{Plots of the temperature dependence of $\chi_{ab}$ and $\chi_c$ under various magnetic fields applied either along or perpendicular to the c-axis. The magnetization measurement was taken after zero-field-cooling the samples first. Robust shielding was observed for $H\parallel c$, indicating strong supercurrent flow in the $ab$-plane. In sharp contrast, the shielding breaks down for mere 1~mT of external magnetic field applied perpendicular to $c$-axis, indicating a weak-link behavior between the superconducting planes ($ab$-planes), a characteristic of Josephson junctions. }
\label{fg:chi} 
\end{figure}

\begin{figure}[t]
\centerline{\includegraphics[width=0.85\columnwidth]{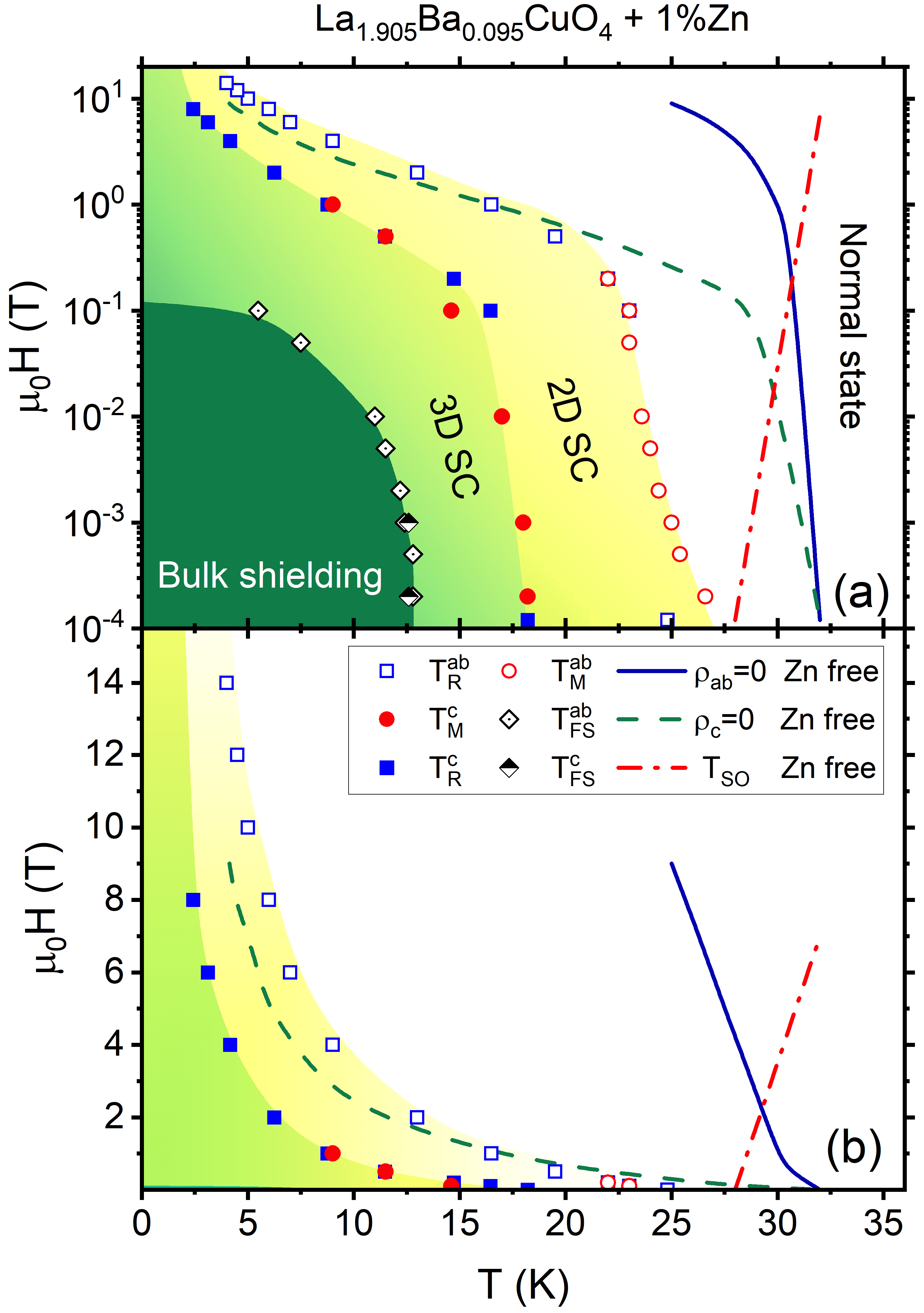}}
\caption{$H$ vs.\ $T$ phase diagram with (a) logarithmic and (b) linear scales for the field. Symbols are defined in the legends; the transition lines for the Zn-free sample are from \cite{wen12b}. }
\label{fg:pd} 
\end{figure}

The dependence of the diamagnetism on the strength of the field is shown in Fig.~\ref{fg:chi}.  Here the range of fields (0.02 mT to 0.5 T) is much smaller than that used in Fig.~\ref{fg:rho} (0.1 to 14 T).  For the case of $\chi_{ab}$, as shown in Fig.~\ref{fg:chi}(a), the transition $T_{\rm FS}^{ab}$ does not shift much at low fields, but once it starts to decrease, the degree of shielding also rapidly decreases.  In the case of $\chi_c$, in Fig.~\ref{fg:chi}(b), the amazing thing is that the degree of shielding decreases much more rapidly than the transition temperature.  Measurements of $M$ vs.\ $H$ indicate that the interlayer superconducting phase stiffness is not sufficient to shield $\mu_0H>2$~mT \cite{SM}.  This loss of shielding occurs even though the resistive transition $T_{\rm R}^c$ is relatively insensitive to the plane-parallel field, as we saw in Fig.~\ref{fg:rho}(d).

We summarize our results with two forms of $H$ vs.\ $T$ phase diagram in Fig.~\ref{fg:pd}, where we consider only ${\bf H}\parallel{\bf c}$.  In order to show the transitions measured by $\chi(T)$, we must plot the field on a log scale in Fig.~\ref{fg:pd}(a).  In this way, we can see that the onset of 2D SC indicated by $T_{\rm M}^{ab}$ is very similar to that determined from resistivity as $T_{\rm R}^{ab}$.  Also, the onset of 3D SC indicated by $T_{\rm M}^c$ is consistent with the resistive $T_{\rm R}^c$.  Of course, the transition towards full bulk shielding, measured by $T_{FS}^{ab}$ and $T_{FS}^c$, occurs at distinctly lower temperature.  This suggests that the initial state of 3D SC achieved on cooling is effectively filamentary.

Plotting the results with a linear field scale, as in Fig.~\ref{fg:pd}(b), makes it more practical to compare with previous results on a Zn-free sample of La$_{1.905}$Ba$_{0.095}$CuO$_4$ \cite{wen12b}.  In the latter case, there is 3D SC transition at 32~K in zero field;  the decoupling of the superconducting layers only appears in finite field.  The onset temperature of 2D SC ($\rho_{ab}=0$, solid line) is relatively insensitive to field, while the onset of 3D SC ($\rho_c=0$, dashed line) decreases rapidly with field before turning up.  With 1\%\ Zn doping, the two transitions are distinct in zero field, and the separation in temperature is relatively constant as the transitions decrease with field.  For fields above $\sim4$~T, the temperatures $T_{\rm R}^{ab}$ and $T_{\rm R}^c$ of the Zn-doped sample seem to parallel the onset of 3D order in the Zn-free sample.

From the $H$-$T$ plot, we see that Zn induces 2D SC correlations in zero magnetic field, {\newr similar to LBCO $x=1/8$ \cite{li07} but} in contrast to the Zn-free {\newr $x=0.095$} sample where this happens {\newr only} at finite field {\newr \cite{wen12b,steg13}}.  Achieving 2D SC but avoiding 3D order requires frustration of the interlayer Josephson coupling.  Such behavior has been rationalized in terms of PDW order, in association with spin and charge stripe orders \cite{berg07,berg09b,agte20}.  It has previously been demonstrated that the Zn enhances the spin-stripe order in LBCO $x=0.095$ \cite{wen12a}, and hence we infer that it also pins and coexists with PDW order.  Related results have recently been reported for Fe-doped LSCO with $x=0.13$ \cite{huan20}.  Now, we have to note that, in contrast to the field-induced state in the Zn-free sample, where the 2D SC state is associated with $\rho_{ab}=0$ \cite{wen12b,steg13}, we always see a finite $\rho_{ab}$ in the 2D SC regime of the Zn-doped sample.  At the same time, we do see %finite (but small) 
{\newr absolute} diamagnetism in $\chi_{ab}$ {\newr with paramagnetic $\chi_{c}$} , so there must be at least small domains of 2D superconducting order within the sample, even if they do not percolate between the voltage contacts on our samples.

%With the present results, we now have evidence that three distinct perturbations of cuprate superconductors (Zn dopants, magnetic vortices, and structural anisotropy) can stabilize both spin-stripe order and substantial 2D SC order without ordering between layers.  In each case, the 2D SC is established over a significant range of temperature.  The common emergent phenomena observed in response to diverse conditions must have a single underlying cause, and that appears to be stabilization of a regime with dominant PDW order.

Zn induces local spin order by frustrating hole motion in its vicinity.  The PDW wave function is expected to have its amplitude go to zero in the middle of a spin stripe \cite{berg07}, so there is no conflict with the impurity presence.  This is in contrast to the uniform $d$-wave state, which empirically does not coexist with local antiferromagnetic order.  Introducing a Zn impurity within a region of uniform superconductivity would require the wave function to decay to zero over the scale of the coherence length.  That large suppression of order might make the PDW order energetically favorable about the impurity.  A similar argument might rationalize the occurrence of PDW order around vortex cores in \bscco\ \cite{edki19}.

While there are similarities between the impact of Zn and magnetic vortices, there are also differences.  The {\newb pancake} vortices tend to form an evenly spaced lattice {\newb within each superconducting layer \cite{clem91,daem93}}, so that if vortices induce PDW order, they will each do it over an equal area.  Interactions between vortex-induced PDW halos might explain the transition to a square vortex lattice observed in LSCO with $x\gtrsim0.15$ for fields above 0.5~T \cite{gila02,chan12c}.  In contrast, the Zn atoms are randomly positioned within the CuO$_2$ planes.  The average area per Zn impurity is $100a^2 = 14.3$~nm$^2$ (where $a=3.78$~\AA\ is the Cu-Cu spacing); the magnetic field required to yield this area per vortex is 145~T, a huge field at which superconductivity would not survive in LBCO.  This comparison makes no sense because the random Zn distribution means that some Zn are close together and some are far apart.  Since Zn must locally pin a spin stripe, closely spaced Zn sites may lead to a frustration of the phase order of the stripes, which would also frustrate PDW order.  At the same time, Poisson statistics indicate that, for 1\%\ Zn doping, the probability of having zero Zn impurities within an area of $100a^2$ is 37\%.  Disorder is a crucial factor here.

The idea of Zn pinning PDW order is also relevant to the interpretation of the gap function determined from the analysis of quasiparticle interference (qpi) measured by STM \cite{wang03,mcel03}.  The qpi is induced by scattering of Bogoliubov quasiparticles from defects such as Zn impurities \cite{pan00a}.  If these defects were to pin PDW order locally, then the qpi likely would be characteristic of the PDW state.  (While we are not aware of evidence for spin-stripe order in Zn-doped \bscco, quasi-static spin stripes induced by Zn doping have been observed in underdoped \ybco\  \cite{such10}.)  Hence, the local superconducting gap would have the form predicted for PDW order, which is gapless along the Fermi arc and opening to a large gap in the antinodal regions \cite{baru08,berg09a}.  Indeed, this is the form of the gap inferred from STM measurements on underdoped \bscco\ samples \cite{kohs08}.  This interpretation resolves the difference in the gap structure from the simple $d$-wave form found in ARPES studies (at least in the near-nodal region) \cite{vish12}.

{\it Acknowledgments}:
We thank S. A. Kivelson for suggesting this experiment and for helpful comments.
This work was supported by the U. S. Department of Energy (DOE), Office of Basic Energy Sciences, Division of Materials Sciences and Engineering, under Contract No.\ DE-SC0012704.

\bibliography{lno,theory,LBCO_Zn}

\newpage

%\includepdf[pages={1,{},{},2,{},3,{},4,{},5,{},6},landscape=false,turn=false]{LBCO_Zn_Supplemental_Material.pdf}

\end{document}